\documentclass{article}
\usepackage{bm}
\usepackage{amsmath}
\usepackage[dvipdfmx]{graphicx}
\usepackage{here}
\usepackage{wrapfig}

\date{}

\renewcommand{\refname}{References}
\begin{document}
\title{Effects of spin fluctuation on the magnetic anisotropy constant of itinerant electron magnets}
\author{\fontsize{8.5pt}{0pt}\selectfont Naofumi Kobayashi, Kazushige Hyodo, and Akimasa Sakuma \\\fontsize{8.5pt}{0pt}\selectfont Department of Applied Physics, Graduate School of Engineering, Tohoku University \\ \fontsize{8.5pt}{0pt}\selectfont (Aoba6-6-05,Aoba-ku, Sendai, 980-8579, Japan)}
\maketitle
\section*{Abstract}
In the disordered local moment picture, we calculated the magnetization ($M$) and magnetic anisotropy energy (MAE) of FePt, CoPt, and MnAl ordered 
alloys and body-centered tegragonal FeCo (bct-FeCo) disordered alloy, assuming spatially fluctuated spin configurations at finite temperatures.  All alloys exhibit the relation 
$K_1(T)/K_1(0)=(M(T)/M(0))^n$ with the exponent $n\approx 2$.  This is consistent with the two-ion anisotropy model, in contrast to the usual single-ion anisotropy model 
exhibiting $n=3$.  
Because these systems have different mechanisms of MAE, we suggest that this relation is a general rule for itinerant electron systems.
\\*
\noindent Key words: magnetic anisotropy constants, itinerant electron magnetism, finite temperature, first principles calculation, 3d transition metals
\section{Introduction}
Recent remarkable developments in information systems require higher density magnetic storage devices.  
However, research and development in this field has faced a serious problem regarding information instability.  
According to numerous studies, memory bits need to satisfy the thermal stability condition given by 
$K_1V/{k_{\rm B}T}\ge 60$, where $K_1$ is the uniaxial (or first-order) anisotropy constant, and
$V$ and $T$ represent the volume of bit and the temperature, respectively. 
Thus, the realization of high density memory by decreasing $V$ is accompanied by instability of this condition.  

Despite this serious issue in the finite-temperature nature of the magnetic anisotropy energy (MAE), few studies were conducted for 3d transition-metal systems before the precise studies of Okamoto et al.\cite{okamoto} and Thiele\cite{thiele} et al. In 2002, these studies experimentally demonstrated that the uniaxial MAE of epitaxial FePt films exhibit the relation of $K_1(T)/K_1(0)=(M(T)/M(0))^{2.1}$ in the low-temperature region. They additionally showed that the Callen--Callen low\cite{callen} $K_1(T)/K_1(0)=(M(T)/M(0))^3$ relation is missing. Shortly thereafter, Skomski et al.\cite{skomski} theoretically showed $K_1(T)/K_1(0)=(M(T)/M(0))^2$ dependence in L10-type CoPt with a classical mean-field two-sublattice model. 

Similarly, Mryasov et al.\cite{mryasov} demonstrated that the $K_1(T)/K_1(0)=(M(T)/M(0))^{2.1}$ relation can be reproduced from the effective two-ion anisotropy term or from anisotropic exchange interactions between Fe spins mediated by induced Pt spins. Mryasov et al.\cite{mryasov} considered the following model Hamiltonian:
\begin{equation}
	H=-\sum_{i,j}J_{i,j}\bm{S}_i\cdot \bm{S}_j-\sum_i D^{(1)}_i(S^z_i)^2-\sum_{\nu} D^{(1)}_{\nu}(m^z_{\nu})^2
	\label{eq:saisho}
\end{equation}
where $\bm{S}_i$ and $\bm{m}_{\nu}$ denote the Fe and Pt spins at the $i$th and $\nu$th site in each sublattice, respectively. 
The first term represents the exchange interaction between Fe spins, and the second and third terms are the single-ion anisotropy terms of Fe and Pt spins, respectively.   The key feature in this model is that the Pt spin moment $\bm{m}_{\nu}$ is induced through the exchange field $\sum_iJ_{{\nu},i}\bm{S}_i$ from the surrounding Fe spins.  Thus, the third term can be rewritten in the form $-\sum_{i,j}D^{(2)}_{i,j}S^z_iS^z_j$ by using the relation 
$\bm{m}_{\nu}=-\chi_{\nu}\sum_iJ_{{\nu},i}\bm{S}_i$. Here, $\chi_{\nu}$ implies the spin susceptibility of Pt, and $D^{(2)}_{i,j}$ is defined by 
$D^{(2)}_{i,j}=\sum_{\nu}D^{(1)}_{\nu}\chi^{2}_{\nu}J_{\nu,i}J_{j,\nu}$. 
This term corresponds to the two-ion anisotropy energy consisting of two different Fe spins. 
 In this sense, the resultant form of the Hamiltonian in eq. (\ref{eq:saisho}) becomes a generalized anisotropic exchange Hamiltonian, 
 which is usually known as the XXZ spin model. Based on this scheme, the $K_1(T)/K_1(0)=(M(T)/M(0))^{2.1}$ 
 relation could be successfully realized for the FePt system using the Langevin dynamics simulation.\cite{nowak}
In 2006, Staunton et al.\cite{staunton} attempted a first-principles calculation for the MAE of FePd at a finite temperature. 
Staunton et al. also demonstrated the relation $K_1(T)/K_1(0)=(M(T)/M(0))^2$  and proposed that the anisotropic exchange interaction is responsible for the 
$K_1(T)/K_1(0)=(M(T)/M(0))^2$ relation. 

According to the above theories, an effective two-ion anisotropy term is realized in the two-sublattice systems: 
one sub-lattice has a strong ferromagnetic character and the other sub-lattice has a large spin-orbit interaction (SOI).  
In the present work, to confirm the above scenario, we calculate the MAE as a function of $M$ for the tetragonally 
distorted ($c/a=1.2$) FeCo disordered alloy\cite{burkert,kota}, which was theoretically proposed by Burkert et al.\cite{burkert} to have large MAE.  
The FeCo disordered alloy is not a two-sublattice system, and it is completely different from the FePt system.  
For reference, we also calculated MAE of FePt, CoPt, and MnAl L10-type ordered alloys.  
We selected MnAl because Al does not have strong SOI like Pt.  
Furthermore, as suggested by Kota et al.,\cite{kota} the mechanisms of MAE occurrence for these systems differs.  
In this sense, the MnAl and FeCo alloys are adequate systems for investigating whether the model proposed by Mryasov et al. is feasible.  Consequently, we observed that these systems exhibit $K_1(T)/K_1(0)=(M(T)/M(0))^n$ relation with the exponent $n$ approximately 2.  
From these results, we expect that the $K_1(T)/K_1(0)=(M(T)/M(0))^n$ ($n\approx 2$) relation is a general feature in ferromagnetic metals.
\section{Calculation model and method}
A standard method for performing practical calculations of the finite-temperature magnetism of itinerant electron systems is to adopt the coherent potential approximation (CPA) in terms of the thermally fluctuated spins as scattering potentials for electrons.\cite{kota} To obtain the MAE, the SOI is further required in the Hamiltonian, as demonstrated by Staunton et al.\cite{staunton} Although the inclusion of SOI requires considerable computer resources and time, this approach is unavoidable for realizing the direct dependence of $M$ and $K_1$ on temperature. 
However, by concentrating only on the relationship between $M(T)$ and $K_1(T)$, this expensive approach can be avoided by employing the disordered local moment (DLM) picture as follows.  First, we assume a certain spin configuration $\left\{\bm{e}_i\right\}$ ($\bm{e}_i$ is a unit vector at the $i$th site) in real space so that the average direction points to $\bm{n}$ (unit vector).  
By artifically arranging the configuration $\left\{\bm{e}_i\right\}$ to vary the summation $(1/N)\sum^N_{i=1}\bm{e}_i$ from 1 to 0, we can realize the states between $T=0$ and $T=T_{\rm C}$ (Curie temperature).  Next, under this configuration ($\bm{n};\left\{\bm{e}_i\right\}$) as a molecular field distribution, we calculate the electronic total energy and magnetization defined by $E(\bm{n};\left\{\bm{e}_i\right\})$ and $M_n\left\{\bm{e}_i\right\}$ 
($n$ component of the magnetization), respectively.  
From the relation between ($E(\bm{n}=\bm{a};\left\{\bm{e}_i\right\})－E(\bm{n}=\bm{c};\left\{\bm{e}_i\right\}))/V$ and $M_n\left\{\bm{e}_i\right\}$, 
we get $K_1(T)$ versus $M(T)$, where $\bm{a}$ and $\bm{c}$ imply the directions of the a- and c-axis, respectively. 
To further reduce computational resources, we assume that the direction $\bm{e}_i$ is restricted to be parallel or antiparallel to $\bm{n}$, 
which means that the spin configuration is always collinear in the line parallel to $\bm{n}$.  
This strongly influences the temperature dependence of $M$ and then $K_1$.  
However, we believe that this assumption does not seriously affect the relationship between $K_1$ and $M$.  

To confirm this model, we calculated the case for FePt and compared our results with those obtained by Mryasov et al.\cite{mryasov} and Staunton et al.\cite{staunton} 
We adopt CPA for the random arrangement of spins $\left\{e^z_i\right\}$.  
Thus, the models we considered here are given as $(\rm{Fe}\uparrow)_{1-X}(\rm{Fe}\downarrow)_XPt$ for FePt (see Fig. 1) and 
$(\rm{Fe}\uparrow)_{1-X}(\rm{Fe}\downarrow)_X(\rm{Co}\uparrow)_{1-X}(\rm{Co}\downarrow)_X$  
for FeCo alloy with $0<X<0.5$, and so on.  Here, $\rm{Fe}\uparrow$ ($\downarrow$) describes an Fe atom whose moment points in the $\bm{n}$ ($－\bm{n}$) direction.  
Therefore, $X=0$ implies a completely ferromagnetic state at $T=0$, and $X=0.5$ implies a non-magnetic state at $T=T_{\rm{C}}$.  
For each $X$ and $\bm{n}$, we calculate both $E(\bm{n};X)$ and $M_n(X)$ by means of the CPA.  
The systems we considered here are FePt, CoPt, MnAl, and FeCo alloys which have large $K_1$ values.  
For the electronic structure calculations, we employed the tight-binding linear muffin-tin orbital (TB-LMTO) method\cite{sakuma} including the SOI under the local density functional approximation.  For the disordered FeCo alloy, we applied another CPA for the random configuration of the Fe and Co atoms in the bct-lattice.
\section{Results and discussion}
In Fig. 2, we show the calculated total magnetic moments $M$ per unit cell of each alloy as a function of $X$.  In every alloy, $M$ decreases linearly with increasing $X$.  This suggests that the magnetic moments in these alloys have similar properties to the local moment model described by the Heisenberg Hamiltonian.  In fact, we confirmed that the amplitude of the local moment in these alloys remains constant in the whole range of $X$.  
Figure 3 shows the calculated values of $K_1$ as a function of $X$.  All $K_1$ values monotonically decrease with an increase of $X$ or a decrease of $M$.  This behavior is understood qualitatively as follows.  Generally, in magnetic systems, the magnetic easy direction and energy anisotropy originate from the connection of the orbital moments with the crystal axis.  Note that the orbital moments are induced by the spin moments through the SOI, and the average orbital moment determines the anisotropy energy rather than the local orbital moment.  Therefore, when the spin moments fluctuate spatially, the average orbital moment decreases, resulting in a decrease of the anisotropy energy.  As we will show in a separate paper, $K_1$ is expressed by the spatial correlation function of orbital moments, which implies that $K_1$ is determined by the expectation value of the orbital moment corresponding to the average orbital moment.

In Fig. 3, we note that the $K_1$ values remain finite at $X=0.5$ corresponding to the Curie temperature.  
This behavior is due to the assumption that restricts the spin moments to be aligned collinearly, from which there remains an energy difference 
between the cases for $n=a$ and $n=c$ even at $X=0.5$.  
This result is incorrect because, at $T=T_{\rm C}$, the spin configuration is completely random, yet the spin configuration is the same for these two cases.  Thus, we consider that the MAE possesses a certain constant value in addition to the $(M(T)/M(0))^k$ term.

By comparing Fig. 2 and Fig. 3, $K_1$ decreases more rapidly than $M$ with increasing temperature, if $X$ is regarded as temperature.  
This implies that the exponent $k$ of $(M(T)/M(0))^k$ is larger than unity.  
Actually, the curves of $K_1$ versus $M$ shown in Fig. 4 clearly exhibit $k$ value larger than unity.  
To determine the exponent, we generate log-log plots as shown in Fig. 5.  The exponents clearly remain around the value of 2.  
We should emphasize that the exponent for FePt is $2.2$, which is almost the same as that given by experiments and theories.  
Apart from the constant value, the relation between $K_1$ and $M$ can be well reproduced by the present approach based on several assumptions.  
In addition, the cases for the MnAl alloy and bct-FeCo disordered alloy also exhibit an exponent of approximately 2.  
As mentioned in the Introduction, these systems are different from the model proposed by Mryasov et al. to introduce the two-ion anisotropy term in the Heisenberg Hamiltonian.  
This leads us to believe that the relation $K_1(T)/K_1(0)=(M(T)/M(0))^n $ ($n\approx 2$) is a general relationship in itinerant electron systems.  
Even though the amplitude of local moments remains constant in the whole range of $X$ (temperature) as shown in Fig. 2, this is apparently different from the case for localized spin systems where the exponent is around 3.\cite{callen}  
As mentioned previously, this difference comes from the difference in the mechanisms: 
$K_1$ of a metallic system is determined through the spatial correlation function of orbital moments while that of a localized 
spin system is expressed by the single-ion anisotropy energy.
\section{Summary}
In the disordered local moment picture, we calculated the magnetization  and magnetic anisotropy energy of FePt, CoPt, and 
MnAl ordered alloys and the bct-FeCo disordered alloy using a first-principles approach combined with CPA. Here, 
we assumed spatially fluctuated spin configurations as a thermal effect at finite temperatures.  
All alloys exhibit a relation $K_1(T)/K_1(0)=(M(T)/M(0))^n$ with the exponent $n \approx 2$.  
This is consistent with the two-ion anisotropy model, in contrast to the case for the single-ion anisotropy model exhibiting $n=3$.  
From the fact that these systems have different mechanisms of MAE, 
we suggest that the relation $K_1(T)/K_1(0)=(M(T)/M(0))^n $ ($n\approx 2$) is general and robust in itinerant electron systems.

\section*{Figure captions}
Fig. 1: Thermal fluctuation of the magnetic moments in this calculation.

\noindent Fig. 2: Calculated magnetic moments as a function of $X$. 

\noindent Fig. 3: Calculated anisotropy constants as a function of $X$. 

\noindent Fig. 4: Calculated anisotropy constants as a function of $M$. 

\noindent Fig. 5: Logarithmic plot of the anisotropy constants as a function of $M$.

\newpage
\begin{figure}[htbp]
	\begin{center}
		\includegraphics[scale=0.5]{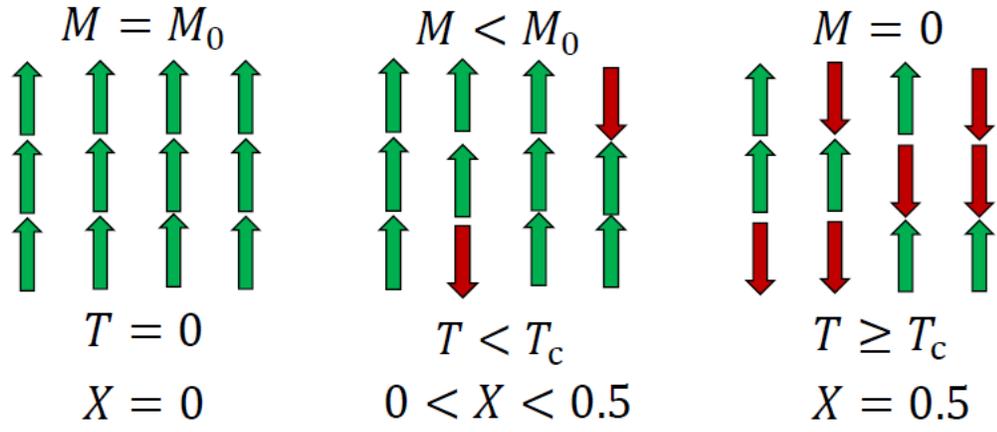}
	\end{center}
	\caption{Thermal fluctuation of the magnetic moments in this calculation. }
\end{figure}
\begin{figure}[htbp]
  \begin{center}
   \includegraphics[width=\hsize]{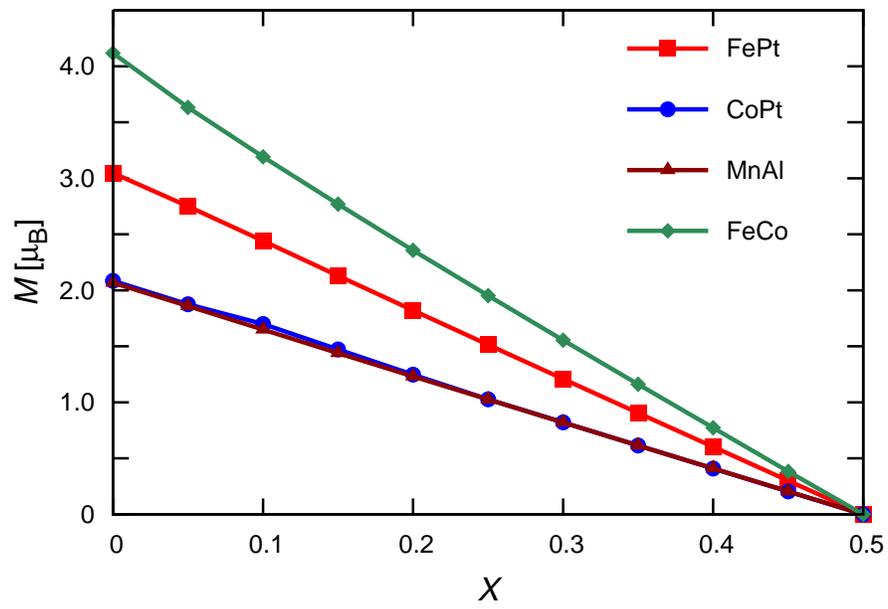}
  \end{center}
  \caption{Calculated magnetic moments as a function of $X$. }
  \label{fig:XM}
\end{figure}
\begin{figure}[htbp]
  \begin{center}
   \includegraphics[width=\hsize]{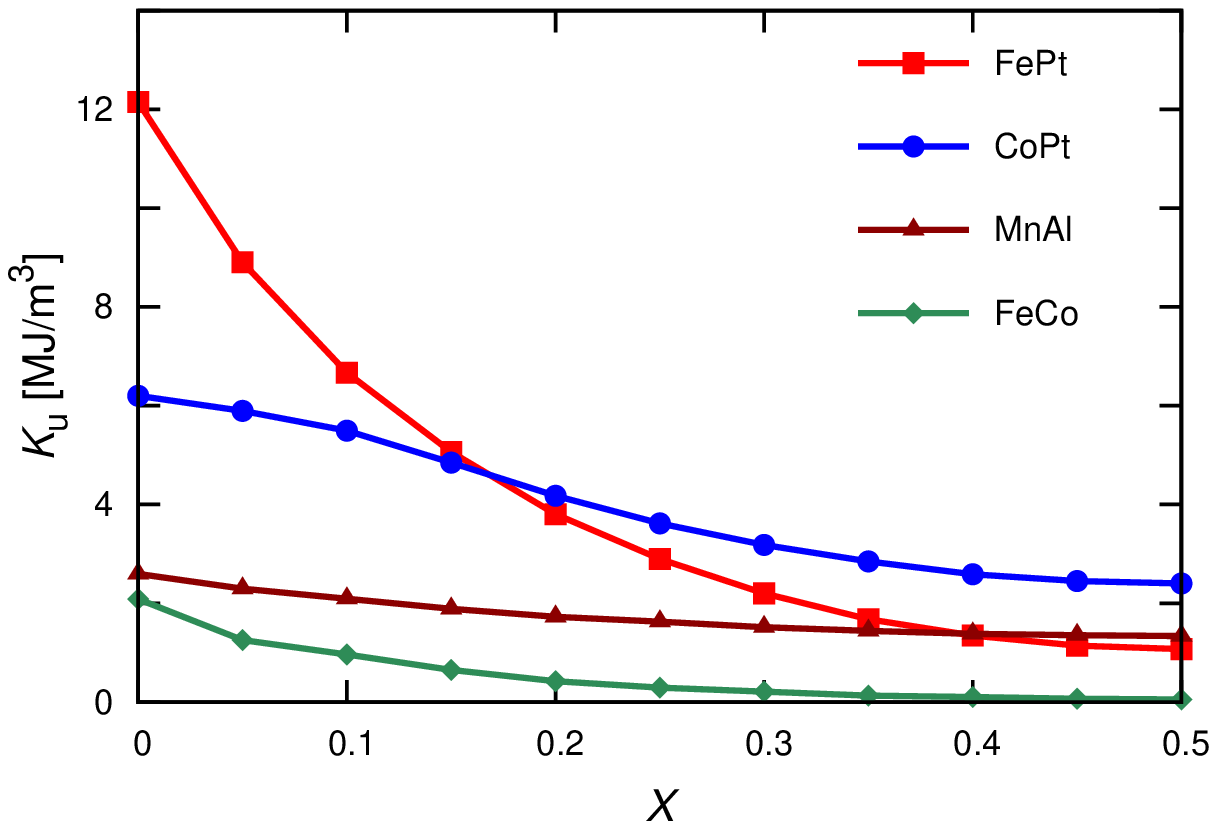}
  \end{center}
  \caption{Calculated anisotropy constants as a function of $X$. }
  \label{fig:XKu}
\end{figure}
\begin{figure}[htbp]
  \begin{center}
   \includegraphics[width=\hsize]{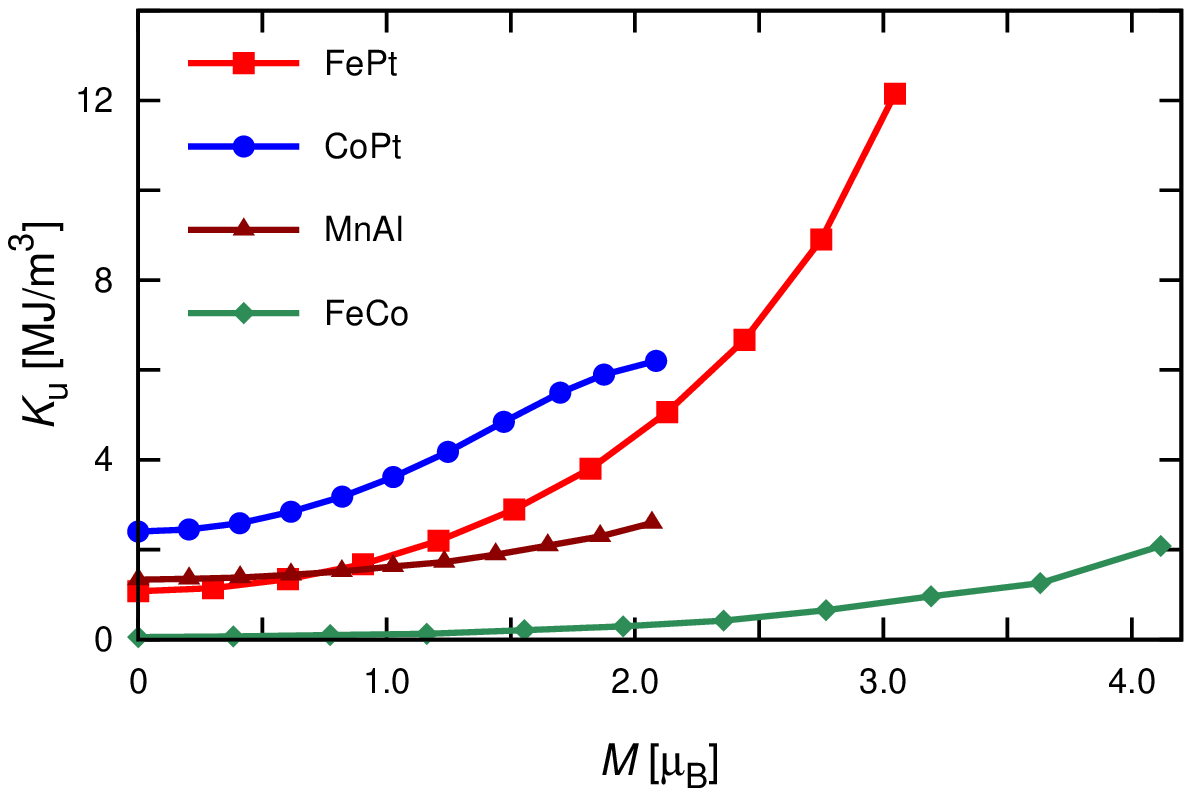}
  \end{center}
  \caption{Calculated anisotropy constants as a function of $M$. }
  \label{fig:MKu}
\end{figure}
\begin{figure}[htbp]
  \begin{center}
   \includegraphics[width=\hsize]{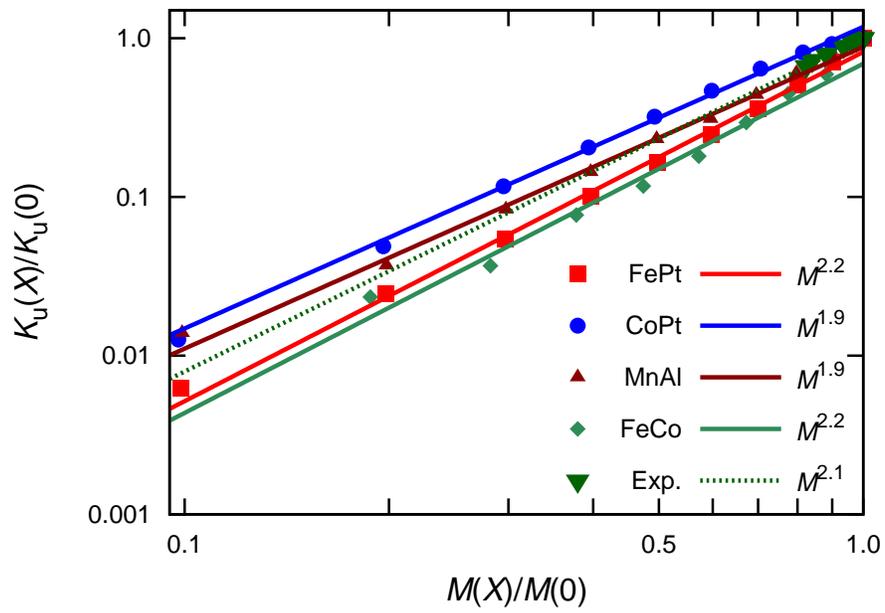}
  \end{center}
  \caption{Logarithmic plot of the anisotropy constants as a function of $M$. }
  \label{fig:log}
\end{figure}
\end{document}